\documentclass[10pt, conference, letterpaper]{IEEEtran}

\usepackage{cite}
\usepackage{graphicx}
\usepackage{epstopdf}
\usepackage{gensymb,bm}
\usepackage[cmex10]{amsmath}
\usepackage{amssymb}
\usepackage{booktabs}
\usepackage{url}
\usepackage{subcaption}

\begin{document}

\title{Double Directional Channel Measurements for THz Communications in an Urban Environment}

\author{\IEEEauthorblockN{Naveed A. Abbasi, Arjun Hariharan, Arun Moni Nair, Ahmed S. Almaiman, Fran\c{c}ois B. Rottenberg,\\ Alan E. Willner, Andreas F. Molisch}
\IEEEauthorblockA{Wireless Devices and Systems Group (WiDeS)\\
	Viterbi School of Engineering\\
	University of Southern California\\
Email: \{nabbasi, arjunhar, moninair, almaiman, frottenb, willner, molisch\}@usc.edu}}

\maketitle
\begin{abstract}
\boldmath
While mm-wave systems are a mainstay for 5G communications, the inexorable increase of data rate requirements and user densities will soon require the exploration of next-generation technologies. Among these, Terahertz (THz) band communication seems to be a promising direction due to availability of large bandwidth in the electromagnetic spectrum in this frequency range, and the ability to exploit its directional nature by directive antennas with small form factors. The first step in the analysis of any communication system is the analysis of the propagation channel, since it determines the fundamental limitations it faces. While THz channels have been explored for indoor, short-distance communications, the channels for {\em wireless access links in outdoor environments} are largely unexplored. In this paper, we present the - to our knowledge - first set of double-directional outdoor propagation channel measurements for the THz band. Specifically, the measurements are done in the 141 - 148.5 GHz range, which is one of the frequency bands recently allocated for THz research by the Federal Communication Commission (FCC). We employ double directional channel sounding using a frequency domain sounding setup based on RF-over-Fiber (RFoF) extensions for measurements over 100 m distance in urban scenarios. An important result is the surprisingly large number of directions (i.e., direction-of-arrival and direction-of-departure pairs) that carry significant energy. More generally, our results suggest fundamental parameters that can be used in future THz Band analysis and implementations.
\end{abstract}


\section{Introduction}
\label{sec:intr}
The demands for higher data rates to cater for applications such as high-resolution videos and 3D experience environments are increasing constantly. In order for the technology to satisfy consumer needs, next-generation wireless networks need to realize data rates reaching terabits per second (Tbps) while also supporting higher connection density \cite{ 5764977, 4623708, 7894280}. The 60 GHz spectrum is expected to allow data rates to tens of Gbits/s \cite{ghasempour2017ieee}, however, the available spectrum around this frequency is limited to 9 GHz. Consequently, it may not fully satisfy the bandwidth requirements for long. To support high throughputs on the order of Tbps, enormous bandwidth is required along with higher area spectral efficiency. Hence, researchers around the world are exploring the Terahertz (THz) Band (0.1 - 10 THz) \cite{7335434, 6882305, 6005345, Kurner2014, Akyildiz201416}: very large swaths of spectrum are either available now, or are expected to become available in the future.

The highly directional nature of THz propagation, combined with the potential of building antenna arrays with a large number of elements ($>$ 1000) that can be fit into a reasonable form factor are key motivations for THz research \cite{kim2014full}. Although THz communication presents an interesting opportunity, a number of challenges are faced by potential communication schemes in this band. Higher frequencies, being highly directional, are more easily blocked by any obstacles on their path. Moreover, atmospheric characteristics also play a major part in communication links operating above 100 GHz, since molecular absorption may lead to high signal losses. We should also note that the ultimate application of a communication scheme is highly dependent on the scenario for its use as well. For instance, large coverage areas, high speed and fast moving networks each pose their own set of challenges. To properly assess the potential and limitations of THz communications, we first need knowledge of propagation channel characteristics, which have to be obtained from measurements. In this regard, recent interest by Federal Communication Commission (FCC) is noticeable where sub-bands in the THz range were earmarked for experimental licenses to encourage research \cite{fcc}. 

Up to now, most of the interest for THz communications has concentrated on very short range indoor scenarios, such as infostations. Consequently, most THz propagation channel measurements have concentrated on this scenarios \cite{chia2010extremely,priebe2011channel,kim2015statisticala,kim2016characterization,khalid2019statistical}. However, THz might also be suitable for outdoor wireless access, in particular for hotspot/femtocell scenarios where the distance between base station (BS) and user equipment (UE) can be up to 100 m. Most outdoor measurement are limited to the pathloss/atmospheric attenuation for line-of-sight (LoS) connections. Campaigns in literature targeting longer distances such as \cite{ma2018invited} focus on narrowband channels. Moreover, these measurements use reflective materials to realize longer distances and therefore they cannot be extended to double-directional channel models that truly represent the long distances. However, to assess wireless access with phased arrays, the double-directional (or MIMO) channel characteristics are required \cite{steinbauer2001double}. 

Keeping the preceding discussion in mind, we present what is to our knowledge the first set of double-directional THz channel measurements in urban street environments for distances above 100 m. To accomplish measurements over such large distances using a frequency domain setup, we used a RF-over-fiber (RFoF) extension of a VNA (vector network analyzer) based system. Double-directional characteristics are obtained by mechanical rotation of horn antennas. Moreover, the measurements were performed in the 141 - 148.5 GHz band, which is one of the widest (7.5 GHz) bands currently allocated by the FCC for THz research \cite{fcc}. 

The rest of this paper is organized as follows. The measurement equipment and sites are described in Section II. Section III highlights the major results for our measurement campaign. Finally, the manuscript is concluded in Section IV.

\section{Measurement Equipment and Site}
\label{sec:definitions}
Experimentation in the THz band is rather difficult due to extremely small wavelengths (1 - 0.1 mm) and the strong signal attenuation by components such as cables that would be unproblematic at lower frequencies. The measurements presented in the current manuscript were conducted with a custom configured frequency domain channel sounder. In this section, we describe the testbed used for the experimentation and provide details of the measurement sites.

\subsection{Testbed Description}
\begin{figure}[t!]
	\centering
	\includegraphics[width=7.5cm]{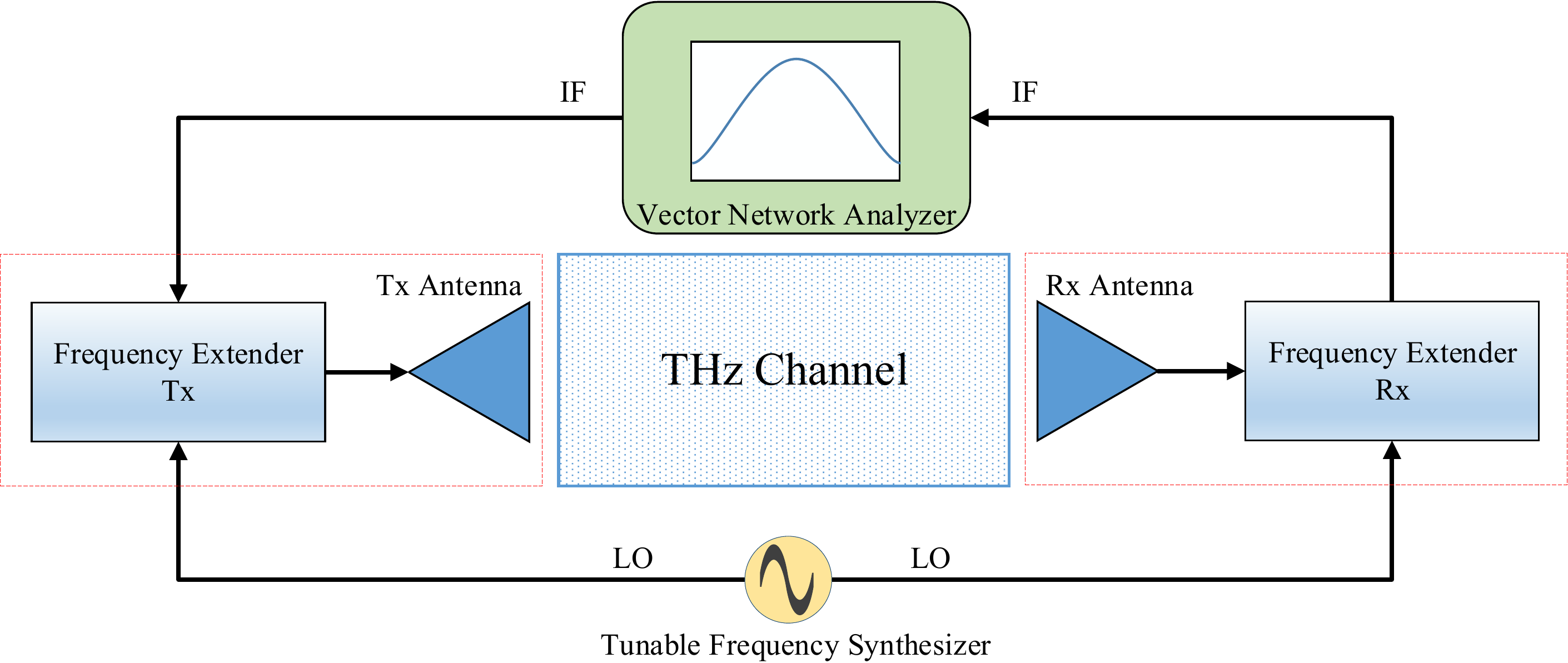}
	\caption{VNA-based THZ measurement setup. \label{fig:Fig1}}
\end{figure}
Measurement of wideband channel characteristics, which are the focus of our interest, can be conducted either by time-domain \cite{xing2018propagation} or frequency-domain \cite{priebe2011channel} setups. As discussed earlier, our current work is based on a frequency-domain setup. Such setups are generally known to provide excellent robustness and often act as ground truths for time-domain setups. 

A diagram of our THz equipment is shown in Fig. \ref{fig:Fig1}. A VNA produces Intermediate Frequency (IF) signals that are then mixed by special frequency extenders with signals from a Local Oscillator (LO) that is operating in the THz range. This THz LO signal is generated by a frequency synthesizer and it is then multiplied up into the THz range in the frequency extenders. In case the VNA has more than 2 ports, it can also produce LO signals required by the setup in Fig. \ref{fig:Fig1}, obviating the need for a separate frequency synthesizer. In our work, we use a 4-port VNA by Keysight Technologies (N5247A) that operates in the 0.01 - 67 GHz range to generate all the LO and the IF signals.
\begin{figure}[t!]
	\centering
	\includegraphics[width=8cm]{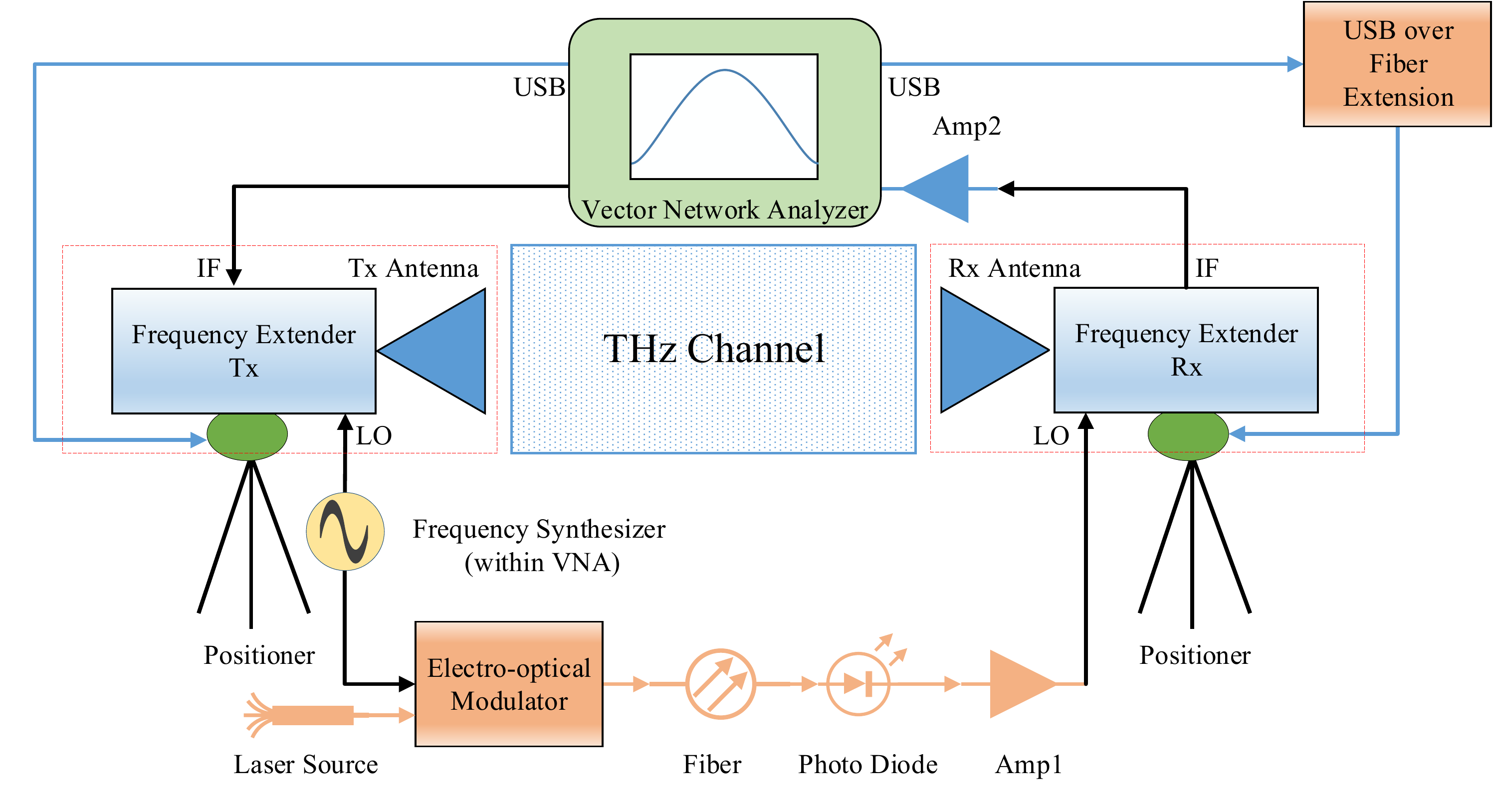}
	\caption{RFoF-based system for THz measurements over longer distances. \label{fig:Fig2}}
\end{figure}
\begin{figure}[b!]
	\centering
	\includegraphics[width=8.3cm]{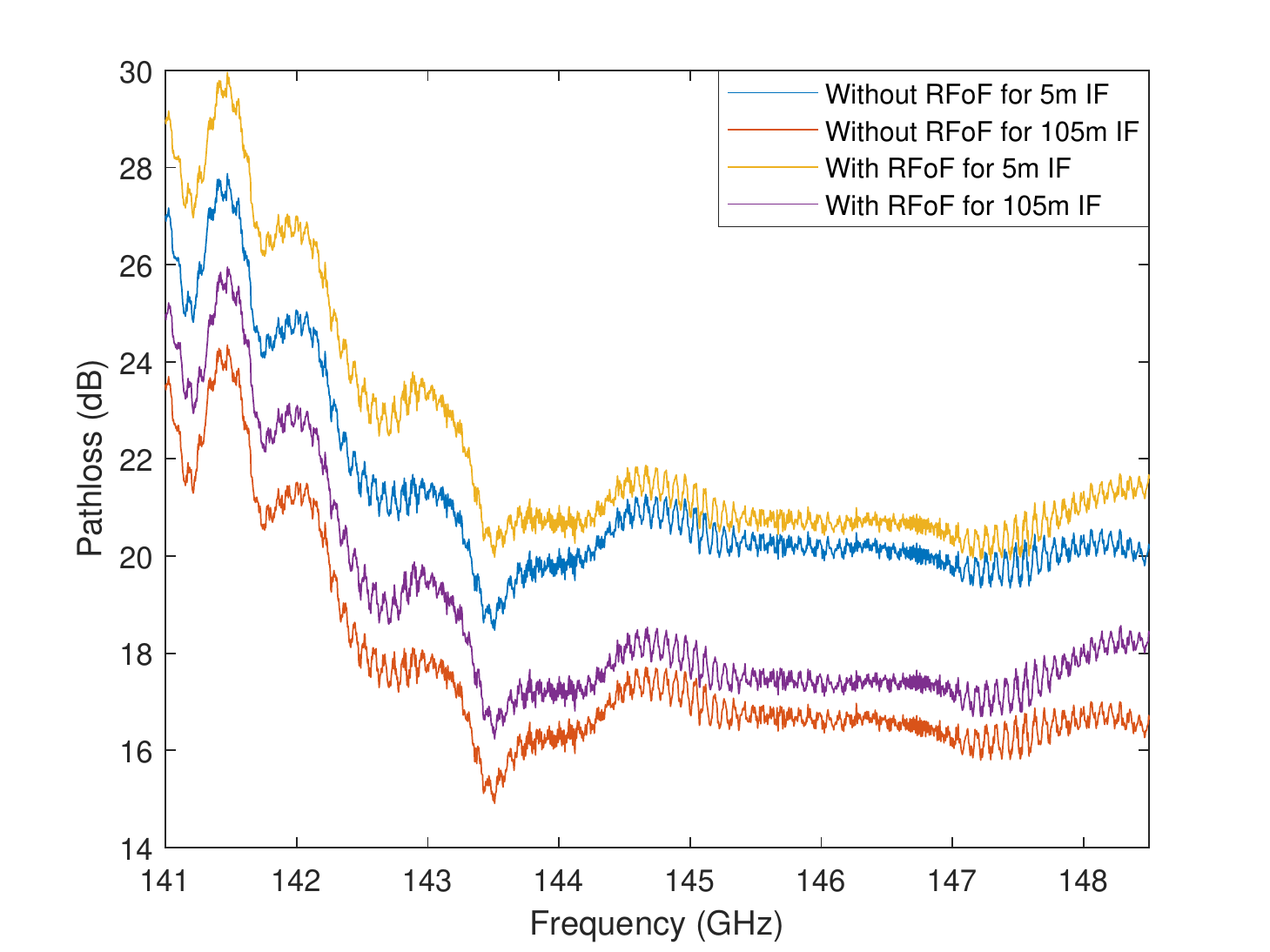}
	\caption{Comparison of pathloss via different configurations of the setup over a 2.4m indoor channel. \label{fig:Fig3}}
\end{figure}

For the frequency extension, we use an extender set produced by Virginia Diodes Inc (VDI) that operates in the range of 140 - 220 GHz range. In order to reach the THz band, the LO signal (11.67 GHz to 18.33 GHz) is multiplied by a factor of 12. The dynamic range of these extenders is on the order of 140 dB for a special high sensitivity mode. It should be noted here that the frequency extenders can have a number of different configurations and though a detailed discussion on the merits of these is beyond the scope of this paper, we make a note that our setup uses the through-reflect configuration since we are mostly interested in transmission characteristics of the channel. The THz sounding signal (mixing product of IF and multiplied LO) is transmitted onto the channel via a directional horn antenna. The reception and downconversion at the receiver is completely analogous. 
\begin{figure*}[t!]
	\centering
	\begin{subfigure}[t]{0.44\textwidth}
		\centering
		\includegraphics[width=8.3cm]{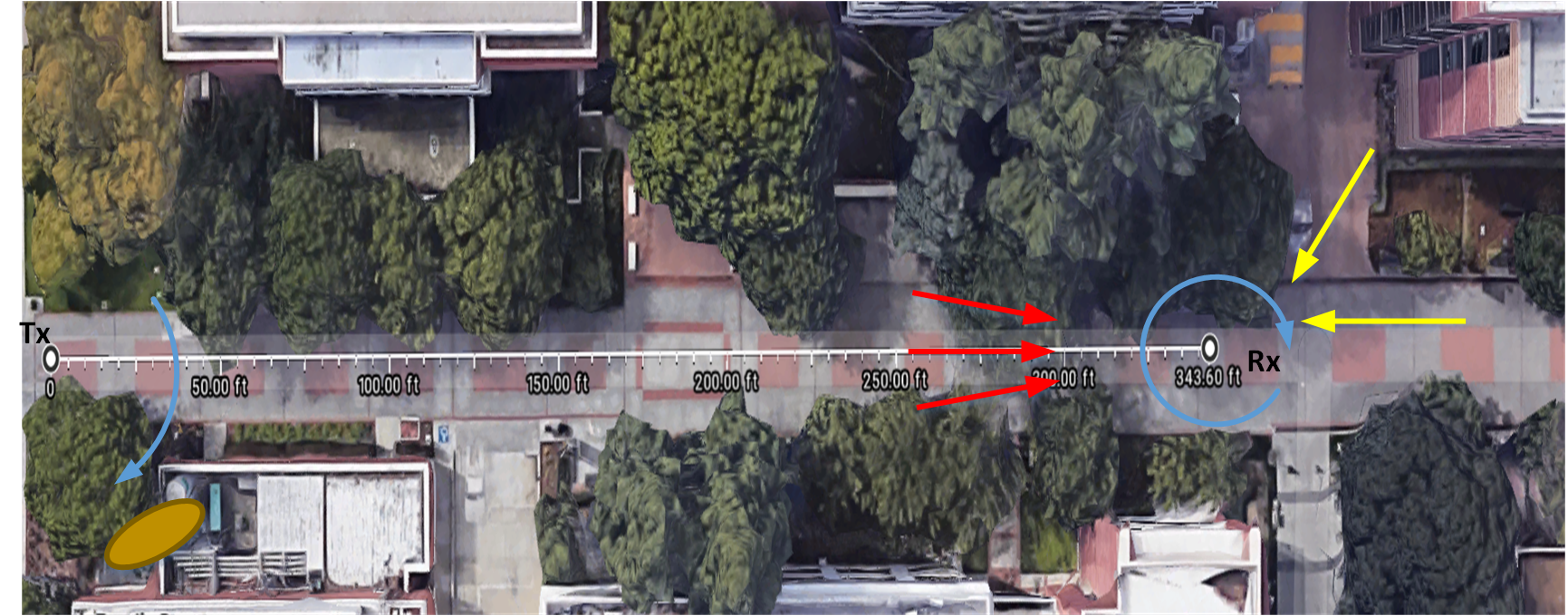}
		\caption{Bloom Walk map.}
	\end{subfigure}~~~
	\begin{subfigure}[t]{0.44\textwidth}
	\centering
	\includegraphics[width=8cm]{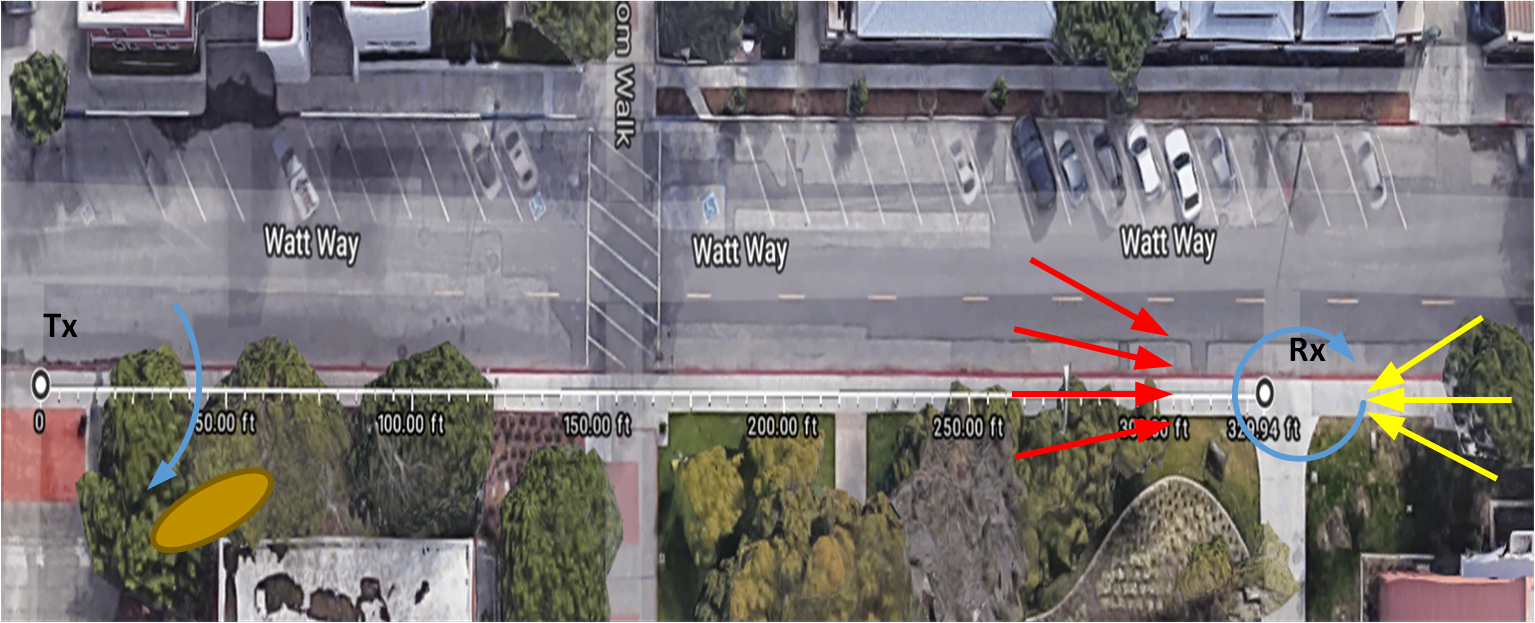}
	\caption{Watt Way map.}
\end{subfigure}
	\caption{Site maps for the two measurement sites.}
	\label{fig:Fig4}
\end{figure*}

Obtaining the {\em double-directional} characteristics of the channel requires us to measure the channel transfer function for different orientations of the transmit and receive horn antennas. Horn rotation is achieved through a rotational positioning system (not shown in Fig. \ref{fig:Fig1}). Since a complete set of double-directional measurements requires a large number of individual measurements, the positioning system needs to be interfaced with the VNA such that the channel measurements can be automated. In our work, we use a positioning system that is capable of performing full 360$^{\circ}$ rotational movements in both azimuth and elevation.

\subsection{RFoF Extension}

A key limitation of a VNA-based THz system stems from the fact that the VNA has to be connected to the frequency extenders via cables and thus, the maximum separation of Tx and Rx extender units (and their antennas) is limited by the cable length. At high frequencies, attenuation and other signal degradation prevent the use of cables beyond a (relatively short) length, thereby limiting the effective channel length (distance between Tx and Rx) that can be explored. For instance, the setup described above uses 5 m cables at both Tx and Rx because of the high cable losses ($\approx$ 2 dB/m) and VNA power limitations. This limits the effective channel length of the setup to 8 - 10 m. Even in case a 10 m distance is achieved, the VNA itself must be in the middle between Tx and Rx,  and can thus act as a major reflector and scatterer. 

To overcome these constraints, we have developed a RFoF solution that can extend the measurable channel length. Fig. \ref{fig:Fig2} shows the design for our RFoF system where we can see that the extension allows the Tx and the VNA to remain together and the Rx unit can extend further than the direct cable limit (5 m for our case). Using an electro-optical modulator, we modulate our RF signal onto a 1550 nm laser. The resulting optical signal is then transmitted over a optical fiber and it is then demodulated using a photo diode. The output of the photo diode is amplified by Amp1 by 33 dB so that it is within the range required by the extender (2 dBm $\pm$ 3 dB). As the Rx is further from the Tx, the IF cable and the USB-based positioner connection need to be extended as well. The IF signal is low frequency (279 MHz), therefore, its cable losses ($\approx$ 0.25 dB/m) are much smaller than those of the LO signal ($\approx$ 2 dB/m) and thus can easily be mitigated by the use of a 25 dB amplifier (Amp2). The positioner USB connection is extended through an off-the-shelf USB-over-fiber extender. Note that though we use 100 m separation currently, larger separations are in principle feasible because fiber attenuation is very low. The main limitations of having longer channel lengths arise from the logistics of handling very long cables. 

In order to quantify the performance of our RFoF design, we tested the setup over an indoor LoS channel with a 2.4 m channel length comparing to a VNA setup without RFoF. The results for this are shown in Fig. \ref{fig:Fig3}. We see that for almost all the cases, the pathloss curve retains its shape. The slight improvement for the case of longer IF cables is due to the higher amplification provided by Amp2.

\subsection{Site Descriptions}
As discussed earlier, scenarios are themselves a very important factor in any channel sounding campaign. For the current measurements, we investigate urban outdoor scenarios. Bloom Walk and Watt Way within the USC campus were selected as the locations of these measurements. Both of these are paved roads surrounded by buildings and trees; Watt Way is wider. Bloom Walk is usually frequented by pedestrian traffic and cyclists whereas the majority of traffic on Watt Way is vehicular with pedestrians usually constrained to the side of the road. 

During the course of the measurement, these street were closed to pedestrian and vehicular traffic to maintain a static environment. We used laser crosshairs and pointers for the alignment of the postioning system after the hardware deployment. An overview of the sites is shown in Fig. \ref{fig:Fig4}. It should be noted here that although the exact distance of the measurements was slightly more than 100 m ($\approx$ 104 m and 100 m), we use the `100 m' word in this manuscript for simplicity of language. In any case, the pathloss difference between these distances is quite small. The blue arcs on Fig. \ref{fig:Fig4} show the direction of positioner movements. 

\section{Measurement Results}
In this section, we present some of our key measurement methodologies and the associated results. 
\subsection{Measurement Parameters}
A summary of key measurement parameters, their acronyms and nominal values is given in Table \ref{table:parameters}. The deployment of the positioners was done such that the angle of 0$^\circ$ for both the Tx and the Rx corresponds to the LoS. The Tx scans a 90$^\circ$ (-45$^\circ$ to 45$^\circ$) sector of the channel with a resolution of 5$^\circ$ in the azimuth. The Rx on the other hand looks at a full 360$^\circ$ (-180$^\circ$ to 180$^\circ$) with a 6$^\circ$ angular resolution. Since the antenna beam width is 13$^\circ$, any resolution below 6.5$^\circ$ is sufficient for a comprehensive channel sounding. It should be noted here that we deliberately kept the elevation of both the Tx and Rx same for the current set of measurements to simplify the analysis, and since in this particular setup (Tx and Rx at the same height) no major variations of the elevation are to be expected. Note that the ground reflection of the LoS falls within the (vertical) beamwidth of the antennas.

\begin{table}[t!]
	\centering
	\caption{Setup parameters.}
	\label{table:parameters}
	\begin{tabular}{|l|l|l|}
		\hline
		\textbf{Parameter}              & \textbf{Symbol}   & \textbf{Value} \\ \hline\hline
		\textit{Measurement points}     & $N$                 & 21           \\
		\textit{Averaging factor}     & $Avg$                 & 10           \\
		\textit{Start frequency}        & $ f_{start} $     & 141 GHz          \\
		\textit{Stop frequency}         & $ f_{stop} $      & 148.5 GHz         \\
		\textit{Bandwidth}              & $BW$              & 7.5 GHz         \\	

		\textit{IF Bandwidth}              & $IF_{BW}$              & 100 Hz         \\				
		\textit{THz IF}                   & $ f_{THz IF} $ & 279 MHz          \\
		\textit{Average noise floor}    & $P_{N}$           & -126.10 dB      \\
		\textit{Dynamic range}          & $DR$         & 146  dB          \\
		\textit{Tx rotation range}   & $Tx_{AZ}$        & [0$^{\circ}$,90$^{\circ}$]           \\ 
		\textit{Tx rotation resolution}   & $\Delta Tx_{AZ}$        & 5$^{\circ}$           \\
		\textit{Rx rotation range}   & $Rx_{AZ}$        & [-180$^{\circ}$,180$^{\circ}$]           \\
		\textit{Tx rotation resolution}   & $Tx_{AZ}$        & 6$^{\circ}$           \\ \hline
	\end{tabular}
\end{table}
\subsection{LoS Measurement} 
\begin{figure}[b!]
	\centering
	\includegraphics[width=8cm]{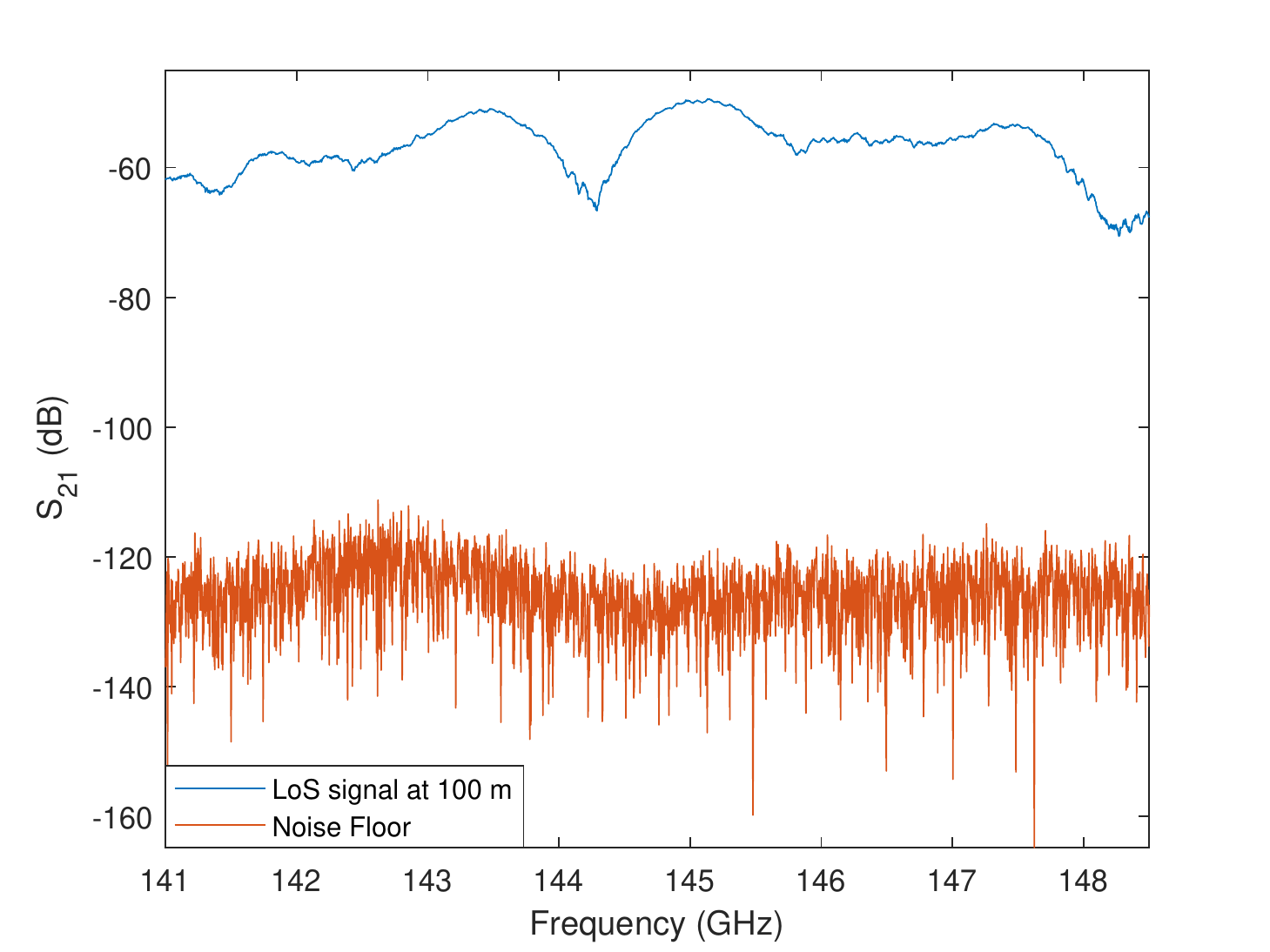}
	\caption{LoS Measurement for the Bloom Walk in comparison with the system noise floor. \label{fig:Fig5}}
\end{figure}
Before moving on to the double directional results, we will look at the specific LoS case for the Bloom Walk measurement. As described earlier, this measurement represents a specific point-to-point communication scenario where the Tx and the Rx are facing each other over a distance beyond 100 m. Fig. \ref{fig:Fig5} shows the results of $S_{21}$ measurements in comparison to the noise floor of the system. Having a significant difference between the two curves ($\approx$ 70 dB) shows that the we have ample $DR$ available for the measurement of multipaths and non-LoS signals using the current setup.
\subsection{Double Directional Measurements}
\begin{figure}[t!]
	\centering
	\begin{subfigure}[t]{0.5\textwidth}
		\centering
		\includegraphics[width=8.3cm]{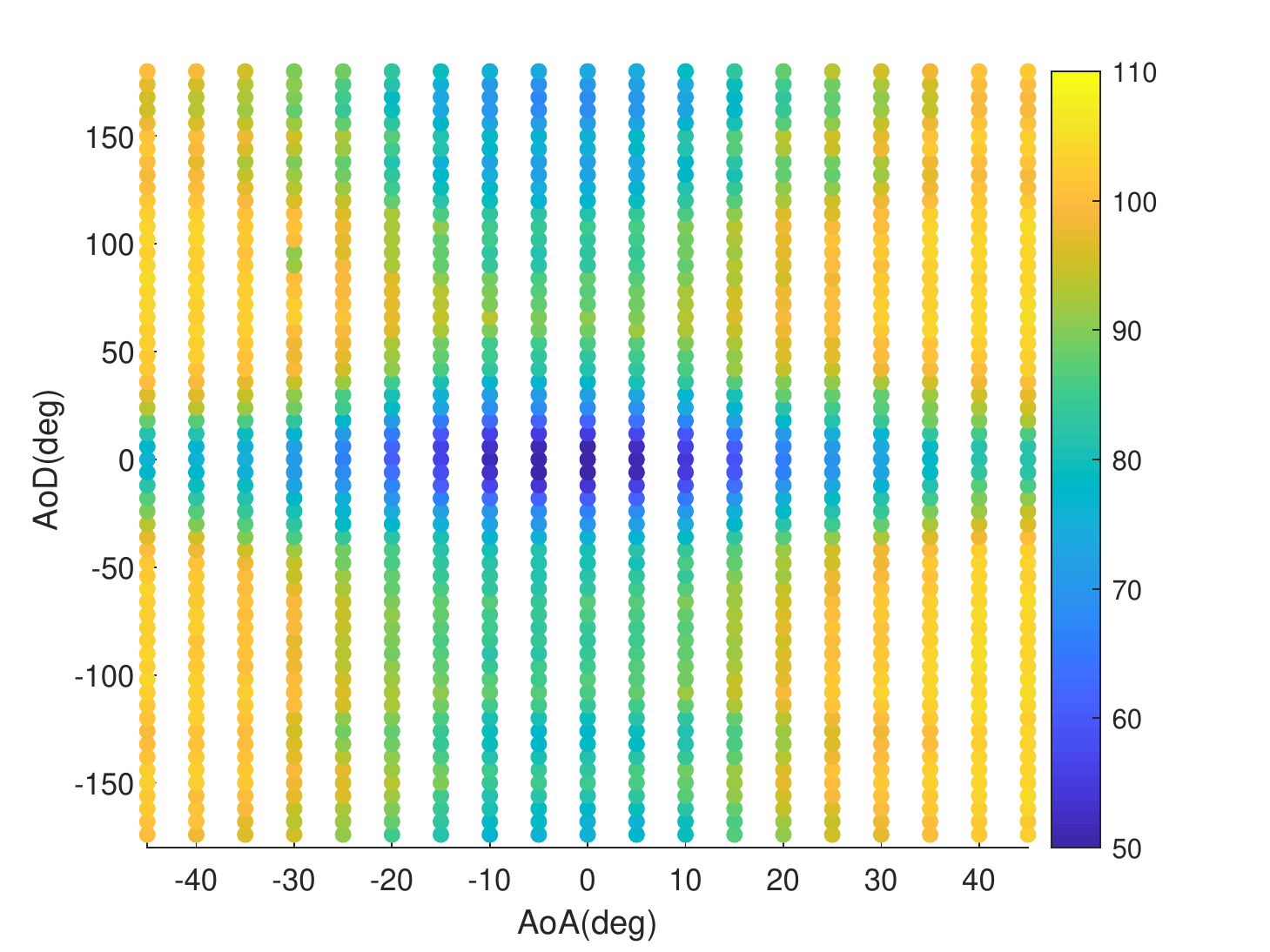}
		\caption{Bloom Walk. \label{fig:Fig7}}
	\end{subfigure}
	
	\begin{subfigure}[t]{0.5\textwidth}
		\centering
		\includegraphics[width=8.3cm]{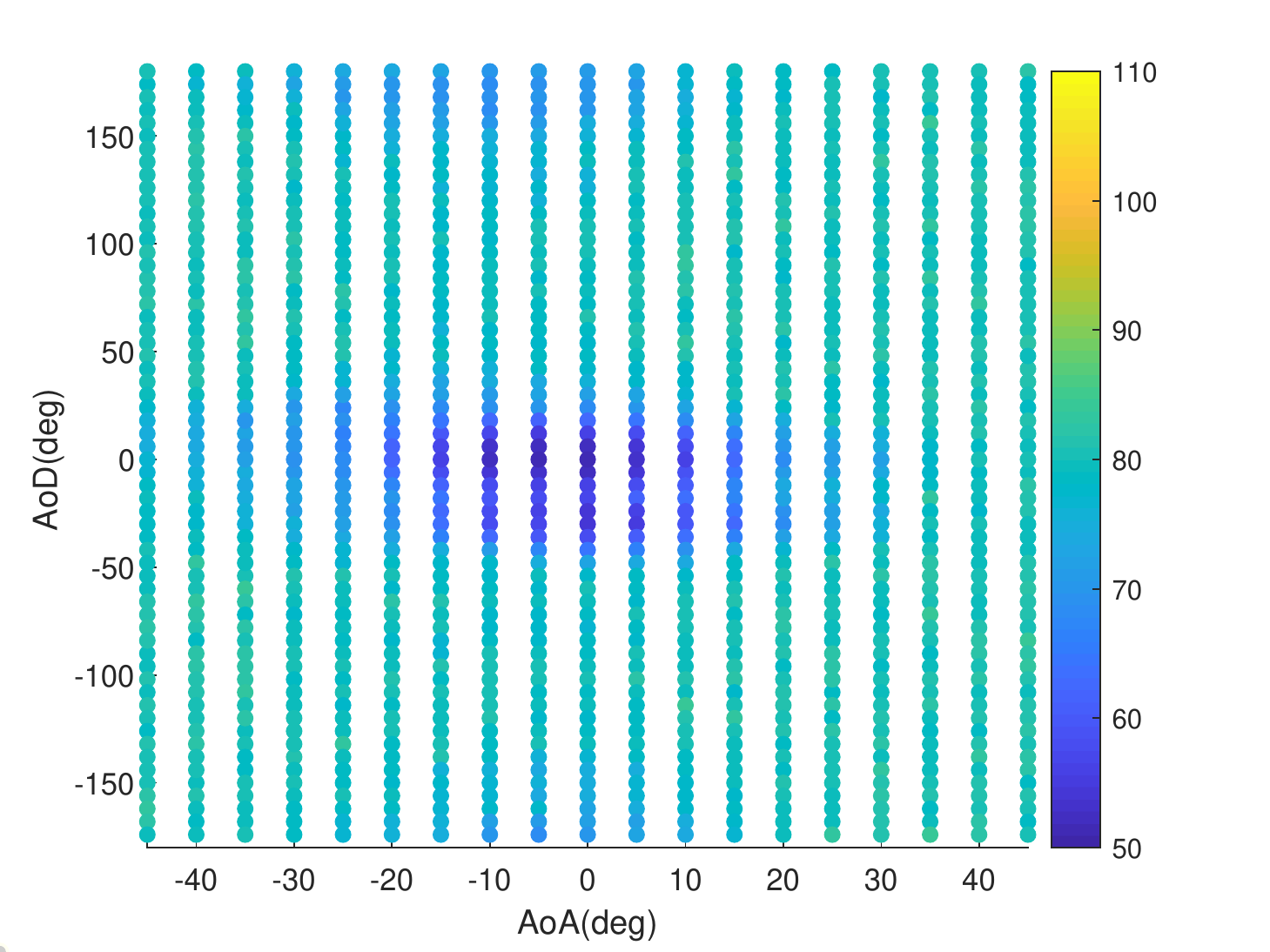}
		\caption{Watt Way. \label{fig:Fig8}}
	\end{subfigure}
	\caption{Average pathloss per direction pair.}
	\label{fig:Fig710}
\end{figure}
Performing channel measurements over the entire angle space of both Tx and Rx with a high value of $N$ and a low value of $IF_{BW}$ is quite time consuming. Averaging the data, which is necessary to reduce the trace noise, takes even longer. In such cases, a complete double directional measurement can sometimes take several days to finish. To keep an outdoor channel static for such long intervals is often not feasible. To solve this issue, we developed a special pre-selection scheme, whose details will be described in \cite{Abbasi_et_al2019}. Since $N$ is low (21) for the current work, a full measurement takes around 3 hours. 

The results for our measurements are shown in Fig. \ref{fig:Fig710} (a) and (b) where we present the average pathloss in each angle bin for different angle of arrival (AoA) and angle of departure (AoD) combinations. We see that for both the scenarios, areas near the LoS suffer only minimum losses. The heavier losses on the right side of the Fig. \ref{fig:Fig710} (a) and (b) covering 30$^\circ$ - 45 $^\circ$ correspond to building walls facing the opposite direction that are identified by brown ellipses in Fig. \ref{fig:Fig4}. Since Watt way is much wider than Bloom walk and one of its sides is not covered by foliage, the angular spread in Fig. \ref{fig:Fig710} (b) seems much larger. It should be noted that for the measurements presented above, despite high losses in several directions, the signal for all direction combinations remains within the $DR$ of the current setup.
\subsection{Angular and Cluster Spreads }
\begin{figure}[t!]
	\centering
	\begin{subfigure}[t]{0.5\textwidth}
		\centering
		\includegraphics[width=8.3cm]{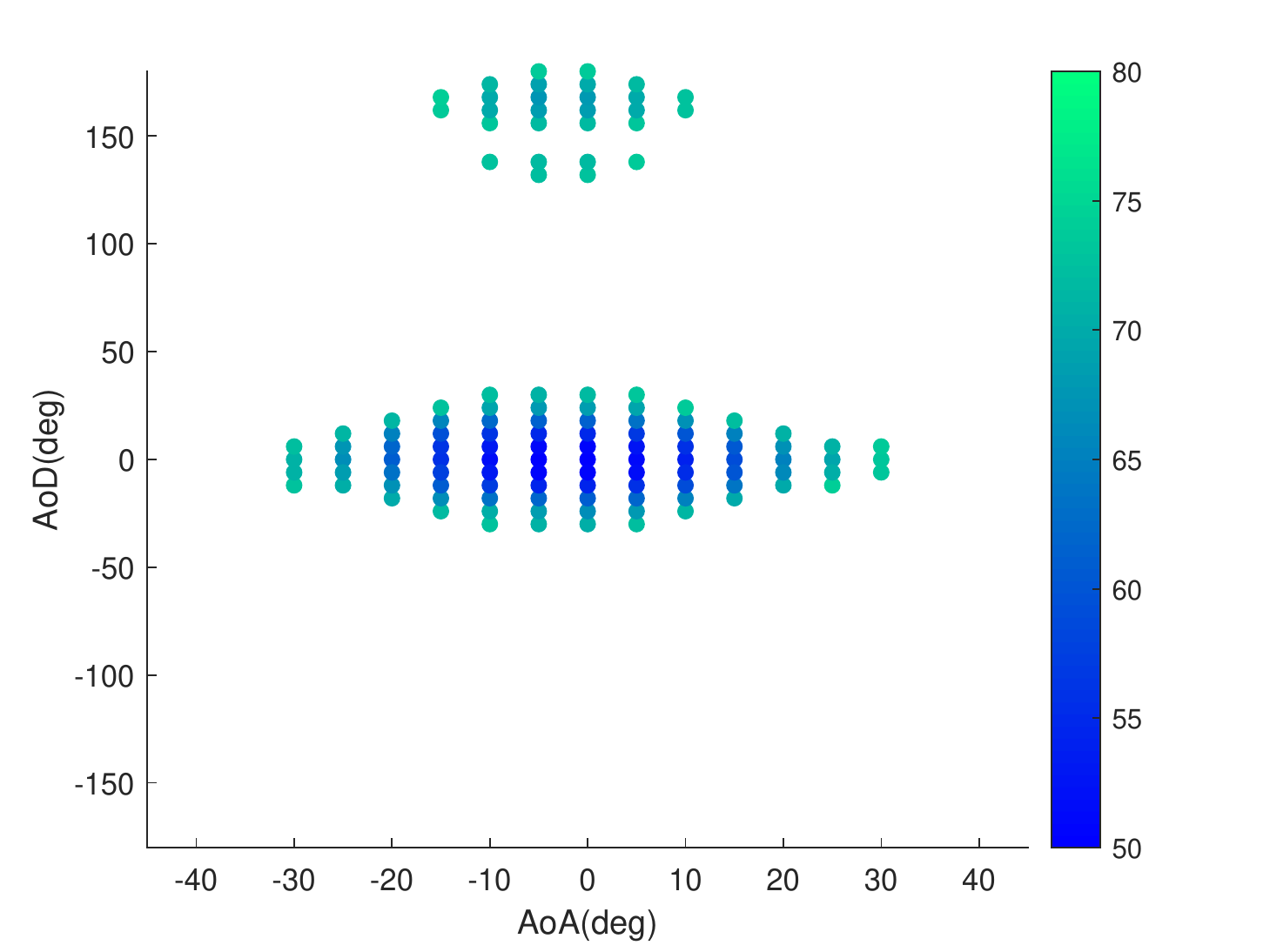}
		\caption{Bloom Walk. \label{fig:Fig9}}
	\end{subfigure}

	\begin{subfigure}[t]{0.5\textwidth}
		\centering
		\includegraphics[width=8.3cm]{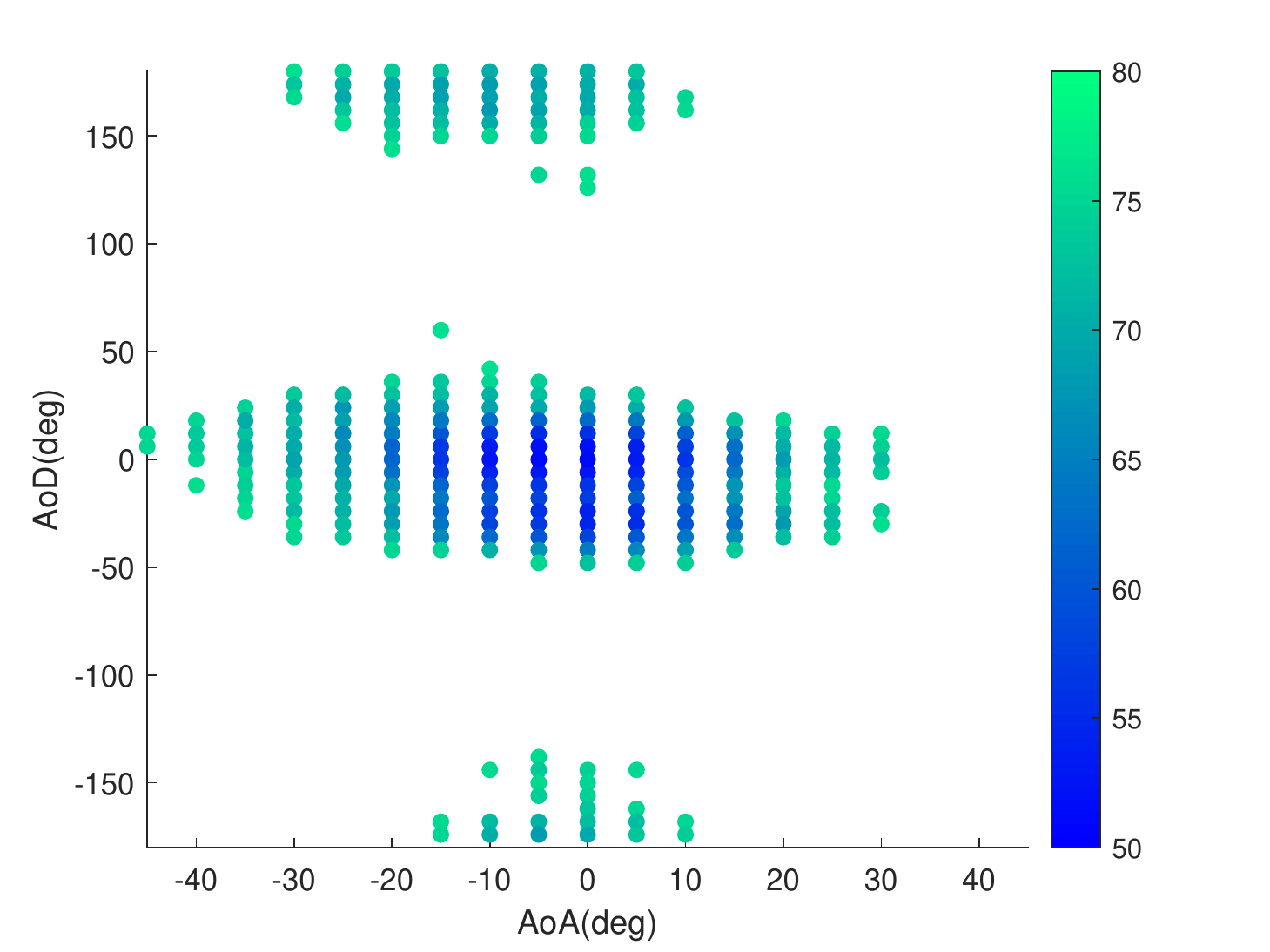}
		\caption{Watt Way. \label{fig:Fig10}}
	\end{subfigure}
	\caption{Low pathloss clusters up to 25 dB above LoS.}
	\label{fig:Fig711}
\end{figure}
RMS (root mean square) azimuth angular spread is defined as \cite{fleury2000first}
\begin{equation}
S = \sqrt{\frac{\int |exp(j\phi)-\mu_\phi|^2 APS(\phi)d\phi}{\int APS(\phi)d\phi}}
\label{eq1}
\end{equation}
where, 
\begin{equation}
\mu_\phi = \frac{\int exp(j\phi) APS(\phi)d\phi}{\int APS(\phi)d\phi}
\end{equation}
This definition has the advantage (compared to the traditional definition as second central moment of the angular power spectrum (APS)) that it is independent of the origin of the spherical coordinate system - in other words, two components at $5^{\circ}$ and $15 ^{\circ}$ result in the same angular spread as two components at $\pm 5 ^{\circ}$. Apart from AoA and AoD RMS angular spreads, another useful measure of channel is the identification of its major clusters and their respective angular spreads. To identify the major clusters for the current measurements, we threshold the results shown in Fig. \ref{fig:Fig710} at 25 dB above the minimum pathloss (LoS measurement) to emulate the DR of a typical sounding system. The results of this are shown in Fig. \ref{fig:Fig711}. We can clearly see that there are two clusters in each case; one around the LoS (C1 hereafter) and one nearly 180$^\circ$ from the LoS (C2 hereafter). The representative physical directions of C1 and C2 on Fig. \ref{fig:Fig4} are marked by red and yellow arrows, respectively. As discussed earlier, the cluster spreads for Watt Way are larger. The exact values of angular and cluster spreads along with average pathloss per cluster are provided in Table \ref{tab:spreads}. The AoA angular spreads for the wider Watt Way is higher by nearly 5$^\circ$. The AoD spreads are much larger since the signal has opportunity to spread over the long channel length. 

For C1, we see that, as expected, cluster spreads are lower for Bloom Walk in comparison to Watt Way, however, the average pathloss is slightly less as well. This results from the fact that the foliage present on both sides of the Bloom Walk attenuates the signal sharply towards the edges of the road leaving only a set of strong components in the middle. Watt Way, having building structures on one side, shows a larger spread and thereby a number of relatively low power components remain on the left side of angular loss spectrum as shown in Fig. \ref{fig:Fig711}. The cluster spreads for C2 follow a similar pattern though it should be noted that the average loss in C2 is much larger since most of the components in this case travel from NLoS directions. The identification of exact paths can be done by means of a power delay profile (PDP) analysis that we intend to perform in future work.
 
\begin{table}[]
	\caption{RMS angular and cluster spreads.}
	\centering
	\label{tab:spreads}
	\begin{tabular}{|l|c|c|}
		\hline
		\textbf{Parameter} & \textbf{Bloom Walk} & \textbf{Watt Way} \\ \hline \hline
		$S_{AoA}$                  & 14.29$^\circ$               & 19.84$^\circ$             \\
		$S_{AoD}$                  & 48.89$^\circ$               & 53.95$^\circ$             \\ \hline 
		$S_{C1,AoA}$                  & 10.90$^\circ$               & 13.24$^\circ$             \\
		$S_{C1,AoD}$                  & 12.33$^\circ$               & 19.04$^\circ$             \\
		Pathloss$_{C1}$ & 60.67 dB & 63.58 dB
		\\ \hline 
		$S_{C2,AoA}$                  & 6.34$^\circ$                & 10.00$^\circ$             \\
		$S_{C2,AoD}$                  & 13.18$^\circ$               & 18.12$^\circ$             \\  Pathloss$_{C2}$ & 71.19 dB & 72.42 dB \\ \hline
	\end{tabular}
\end{table}

\begin{figure}[t!]
	\centering
	\begin{subfigure}[t]{0.5\textwidth}
		\centering
		\includegraphics[width=8.3cm]{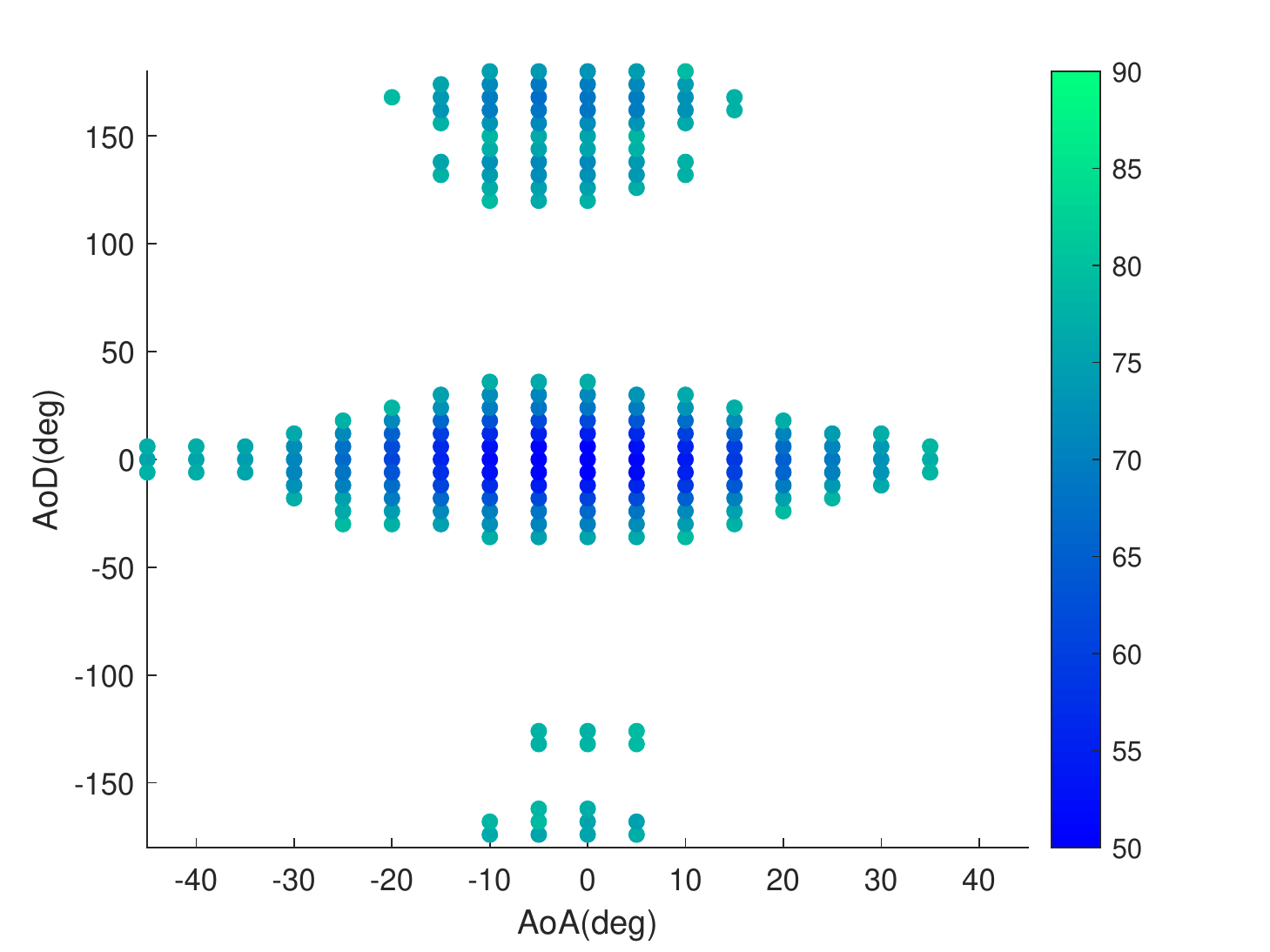}
		\caption{Bloom Walk.}
	\end{subfigure}
	
	\begin{subfigure}[t]{0.5\textwidth}
		\centering
		\includegraphics[width=8.3cm]{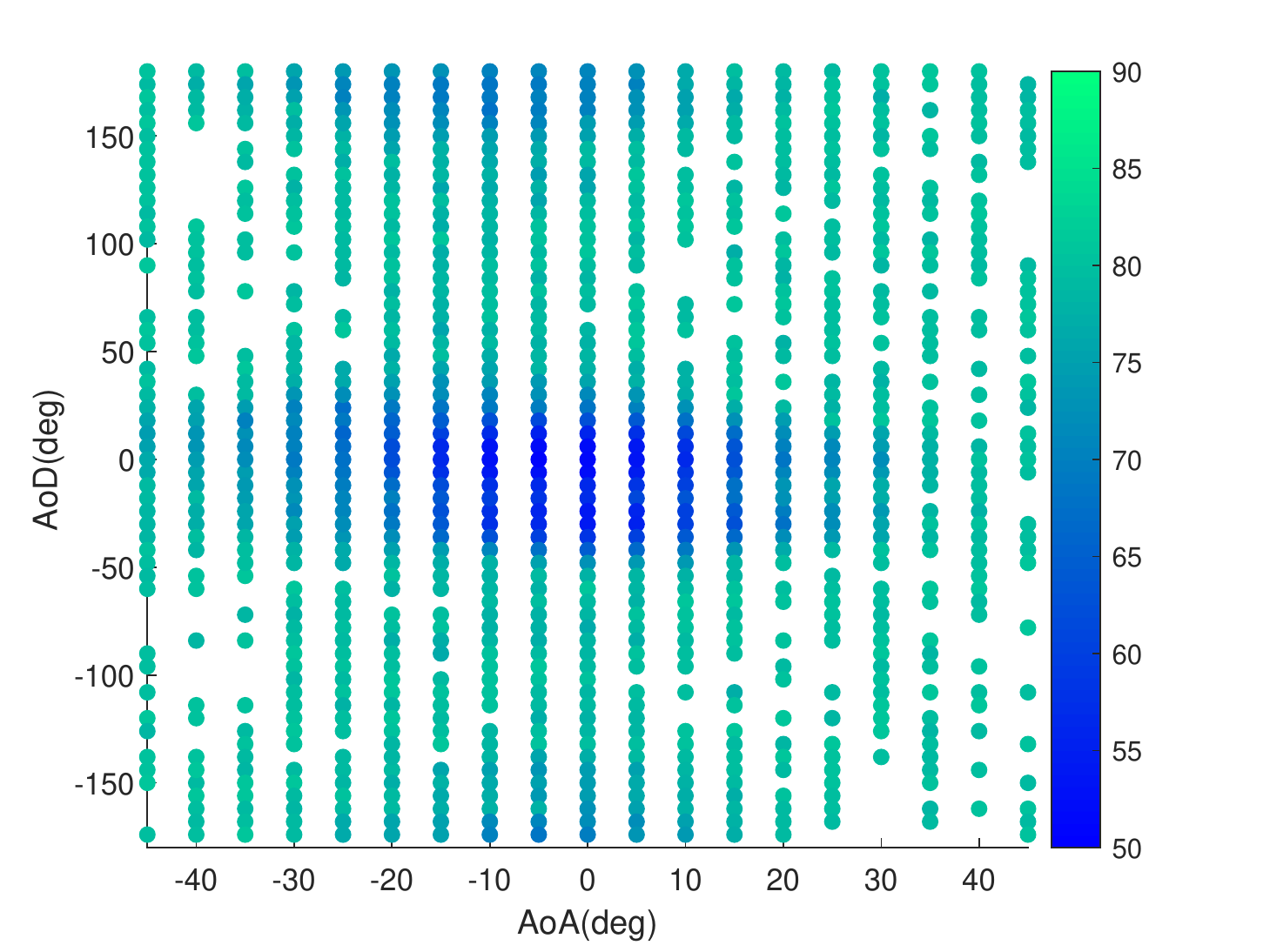}
		\caption{Watt Way.}
	\end{subfigure}
	\caption{Low pathloss clusters up to 30 dB above LoS.}
	\label{fig:Fig12}
\end{figure}
Although the 25 dB threshold in Fig. \ref{fig:Fig711} represents the DR of a typical channel sounder, it is pertinent for us to note the richness of diffuse components in the channel. To elaborate on this, instead of a 25 dB threshold, in Fig. \ref{fig:Fig12}, we use a 30 dB threshold. We can clearly see that the number of direction pairs with significant energy has increased by a large number especially for Watt Way where almost all the direction pairs are below the pathloss threshold. We stress that at least some of these components cannot be explained as ``sidelobes" of the LOS component: since the sidelobe suppression in our system is 25 dB, those angle combinations where neither TX nor RX is pointed towards the LOS are suppressed by 50 dB. Thus, the observed 30 dB components in the observed range must correspond to physically existing multipath components. 
Such rich, low-power components might be well modeled as ``diffuse background" similar to the signal model of \cite{richter2005estimation}. They are especially relevant for multi-user systems, where seemingly weak components from one user might still represent significant interference to another user (near-far effect).  
\section{Conclusion}
\label{sec:conclusion}
In this paper, we present the first set of double-directional outdoor THz channel measurements performed over distance of 100 m in the range of 141 - 148.5 GHz band for urban channels. In order to achieve these measurements, we developed a novel RFoF-based extension for our frequency domain channel sounder. Our results show relatively low pathloss values for outdoor THz channels in urban environments. In our future work, we plan to investigate further indoor and outdoor scenarios, as well as multiple-input multiple-output (MIMO) channels.

\section{Acknowledgment}
This work was supported by the Semiconductor Research Corporation (SRC) under the ComSenTer program. We thank  Dr. Todd Younkin (SRC) and Prof. Mark Rodwell (UCSB) for helpful discussions. 	

\small
\bibliographystyle{IEEEtran}
\bibliography{IEEEabrv,bibliography}

\begin{thebibliography}{10}
\providecommand{\url}[1]{#1}
\csname url@samestyle\endcsname
\providecommand{\newblock}{\relax}
\providecommand{\bibinfo}[2]{#2}
\providecommand{\BIBentrySTDinterwordspacing}{\spaceskip=0pt\relax}
\providecommand{\BIBentryALTinterwordstretchfactor}{4}
\providecommand{\BIBentryALTinterwordspacing}{\spaceskip=\fontdimen2\font plus
\BIBentryALTinterwordstretchfactor\fontdimen3\font minus
  \fontdimen4\font\relax}
\providecommand{\BIBforeignlanguage}[2]{{%
\expandafter\ifx\csname l@#1\endcsname\relax
\typeout{** WARNING: IEEEtran.bst: No hyphenation pattern has been}%
\typeout{** loaded for the language `#1'. Using the pattern for}%
\typeout{** the default language instead.}%
\else
\language=\csname l@#1\endcsname
\fi
#2}}
\providecommand{\BIBdecl}{\relax}
\BIBdecl

\bibitem{5764977}
K.-C. Huang and Z.~Wang, ``{Terahertz Terabit Wireless Communication},''
  \emph{Microwave Magazine, IEEE}, vol.~12, no.~4, pp. 108--116, June 2011.

\bibitem{4623708}
V.~Chandrasekhar, J.~G. Andrews, and A.~Gatherer, ``Femtocell networks: a
  survey,'' \emph{IEEE Communications Magazine}, vol.~46, no.~9, pp. 59--67,
  September 2008.

\bibitem{7894280}
M.~Shafi, A.~F. Molisch, P.~J. Smith, T.~Haustein, P.~Zhu, P.~D. Silva,
  F.~Tufvesson, A.~Benjebbour, and G.~Wunder, ``{5G: A Tutorial Overview of
  Standards, Trials, Challenges, Deployment and Practice},'' \emph{IEEE Journal
  on Selected Areas in Communications}, 2017.

\bibitem{ghasempour2017ieee}
Y.~Ghasempour, C.~R. da~Silva, C.~Cordeiro, and E.~W. Knightly, ``Ieee 802.11
  ay: Next-generation 60 ghz communication for 100 gb/s wi-fi,'' \emph{IEEE
  Communications Magazine}, vol.~55, no.~12, pp. 186--192, 2017.

\bibitem{7335434}
A.~Hirata and M.~Yaita, ``{Ultrafast Terahertz Wireless Communications
  Technologies},'' \emph{IEEE Transactions on Terahertz Science and
  Technology}, vol.~5, no.~6, pp. 1128--1132, Nov 2015.

\bibitem{6882305}
I.~F. Akyildiz, J.~M. Jornet, and C.~Han, ``{TeraNets: ultra-broadband
  communication networks in the terahertz band},'' \emph{IEEE Wireless
  Communications}, vol.~21, no.~4, pp. 130--135, August 2014.

\bibitem{6005345}
H.~J. Song and T.~Nagatsuma, ``{Present and Future of Terahertz
  Communications},'' \emph{IEEE Transactions on Terahertz Science and
  Technology}, vol.~1, no.~1, pp. 256--263, Sept 2011.

\bibitem{Kurner2014}
T.~K{\"u}rner and S.~Priebe, ``{Towards THz Communications - Status in
  Research, Standardization and Regulation},'' \emph{Journal of Infrared,
  Millimeter, and Terahertz Waves}, vol.~35, no.~1, pp. 53--62, 2014.

\bibitem{Akyildiz201416}
I.~F. Akyildiz, J.~M. Jornet, and C.~Han, ``{Terahertz band: Next frontier for
  wireless communications },'' \emph{Physical Communication}, vol.~12, pp. 16
  -- 32, 2014.

\bibitem{kim2014full}
Y.~Kim, H.~Ji, J.~Lee, Y.-H. Nam, B.~L. Ng, I.~Tzanidis, Y.~Li, and J.~Zhang,
  ``Full dimension mimo (fd-mimo): The next evolution of mimo in lte systems,''
  \emph{IEEE Wireless Communications}, vol.~21, no.~2, pp. 26--33, 2014.

\bibitem{fcc}
FCC, ``Fcc takes steps to open spectrum horizons for new services and
  technologies,''
  \url{http://https://docs.fcc.gov/public/attachments/DOC-356588A1.pdf}, 2019.

\bibitem{chia2010extremely}
M.~Y.-W. Chia, B.~Luo, and C.~K. Ang, ``Extremely wideband multipath
  propagation channel from 285 to 325 ghz for a typical desk-top environment,''
  in \emph{Infrared Millimeter and Terahertz Waves (IRMMW-THz), 2010 35th
  International Conference on}.\hskip 1em plus 0.5em minus 0.4em\relax IEEE,
  2010, pp. 1--1.

\bibitem{priebe2011channel}
S.~Priebe, C.~Jastrow, M.~Jacob, T.~Kleine-Ostmann, T.~Schrader, and
  T.~K{\"u}rner, ``{Channel and propagation measurements at 300 GHz},''
  \emph{Antennas and Propagation, IEEE Transactions on}, vol.~59, no.~5, pp.
  1688--1698, 2011.

\bibitem{kim2015statisticala}
S.~Kim and A.~G. Zaji{\'c}, ``Statistical characterization of 300-ghz
  propagation on a desktop,'' \emph{IEEE Transactions on Vehicular Technology},
  vol.~64, no.~8, pp. 3330--3338, 2015.

\bibitem{kim2016characterization}
S.~Kim and A.~Zaji{\'c}, ``Characterization of 300-ghz wireless channel on a
  computer motherboard,'' \emph{IEEE Transactions on Antennas and Propagation},
  vol.~64, no.~12, pp. 5411--5423, 2016.

\bibitem{khalid2019statistical}
N.~Khalid, N.~A. Abbasi, and O.~B. Akan, ``Statistical characterization and
  analysis of low-thz communication channel for 5g internet of things,''
  \emph{Nano Communication Networks}, p. 100258, 2019.

\bibitem{ma2018invited}
J.~Ma, R.~Shrestha, L.~Moeller, and D.~M. Mittleman, ``Invited article: Channel
  performance for indoor and outdoor terahertz wireless links,'' \emph{APL
  Photonics}, vol.~3, no.~5, p. 051601, 2018.

\bibitem{steinbauer2001double}
M.~Steinbauer, A.~F. Molisch, and E.~Bonek, ``The double-directional radio
  channel,'' \emph{IEEE Antennas and propagation Magazine}, vol.~43, no.~4, pp.
  51--63, 2001.

\bibitem{xing2018propagation}
Y.~Xing and T.~S. Rappaport, ``Propagation measurement system and approach at
  140 ghz-moving to 6g and above 100 ghz,'' in \emph{2018 IEEE Global
  Communications Conference (GLOBECOM)}.\hskip 1em plus 0.5em minus 0.4em\relax
  IEEE, 2018, pp. 1--6.

\bibitem{Abbasi_et_al2019}
N.~A. Abbasi~et al., ``Double-directional channel measurement in the thz
  band,'' \emph{to be submitted}, 2019.

\bibitem{fleury2000first}
B.~H. Fleury, ``First-and second-order characterization of direction dispersion
  and space selectivity in the radio channel,'' \emph{IEEE Transactions on
  Information Theory}, vol.~46, no.~6, pp. 2027--2044, 2000.

\bibitem{richter2005estimation}
A.~Richter, ``Estimation of radio channel parameters: Models and
  algorithms.''\hskip 1em plus 0.5em minus 0.4em\relax ISLE, 2005.

\end{thebibliography}

\end{document}